\documentclass[number, 5p]{elsarticle}
\biboptions{merge,elide}
\usepackage[T1]{fontenc}
\usepackage[utf8]{inputenc}
\usepackage[english]{babel}
\usepackage{lmodern, amsmath, amsfonts, cancel, graphicx}
\usepackage[colorlinks=true, citecolor=blue, urlcolor=blue, linkcolor=blue, breaklinks=true]{hyperref}

\newcommand*{\niente}[1]{#1}
\newcommand*{\nlfc}{NLFC}
\newcommand*{\blv}{BLV}
\newcommand*{\np}{\ensuremath{^\text{\tiny NP}}}
\newcommand*{\LQ}{\ensuremath{^\text{\tiny LQ}}}
\newcommand*{\RPV}{\ensuremath{^\text{\tiny RPV}}}
\newcommand*{\fb}{\ensuremath{~\text{fb}}}
\newcommand*{\gev}{\ensuremath{~\text{GeV}}}

\begin{document}
\author[ucl]{Gauthier Durieux}
\author[ucl]{Jean-Marc Gérard}
\author[ucl]{Fabio Maltoni}
\address[ucl]{Centre for Cosmology, Particle Physics and Phenomenology (CP3), Université catholique de Louvain, Chemin du Cyclotron 2, B-1348 Louvain-la-Neuve, Belgium}

\author[lpsc,ipnl]{Christopher Smith}
\address[lpsc]{LPSC, Université Joseph Fourier Grenoble 1, CNRS/IN2P3, Institut Polytechnique de
Grenoble, rue des Martyrs 53, 38026 Grenoble Cedex, France.}
\address[ipnl]{IPNL, Université Lyon 1, CNRS/IN2P3, 69622 Villeurbanne Cedex, France.}
\date{\today}
\title{Three-generation baryon and lepton number violation at the LHC}
\begin{abstract}
One of the most puzzling questions in particle physics concerns the status of the baryon ($B$) and lepton ($L$) quantum numbers. On the theoretical side, most new physics scenarios naturally lead to their non-conservation and some amount of violation is actually needed to explain the baryon asymmetry of the Universe. On the experimental side, low-energy constraints such as those on proton decay are so stringent that it is generally believed that no $B$ and $L$ violation will ever be seen in laboratories. We observe that this apparent contradiction, however, disappears when the flavor symmetries involving all three generations are taken into account. We then identify model-independent classes of $B$ and/or $L$ violating six-fermion-based processes that indeed simultaneously satisfy low-energy constraints and produce clearly identifiable signals at the LHC. Finally, through simplified models, we study two classes characterized by $(\Delta B;\Delta L) = (\pm 1;\pm 3)$ and $(\pm2;0)$, that lead to particularly striking signatures ($t \,\mu^+e^+$ and $\bar t\, \bar t\,+$jets, respectively).
\\
\end{abstract}
% PACS:
% Symmetry, in theory of fields and particles, 11.30.-j
% Conservation laws, fields and particles, 11.30.-j
% Baryon number, 11.30.Fs
% Lepton number, 11.30.Fs
% Flavor symmetries, 11.30.Hv
% Unified field theories, models beyond the standard models, 12.60.-i
% Quarks, top quarks, 14.65.Ha
%
\maketitle
\section{Introduction}
The Standard Model (SM), while now experimentally confirmed in all its predictions, cannot be the ultimate theory. Reasons are multifold, and range from mostly aesthetic theoretical arguments (e.g., naturalness, unification of gauge couplings, inclusion of gravity) to experimental ones (e.g., neutrino masses, evidence for dark matter). Moreover, when analyzed in more detail, both theory and data often point toward the existence of new degrees of freedom around the TeV scale, which may soon reveal themselves at the LHC. In this context, the tiny neutrino masses and the extremely long lifetime of the proton remain some of the most puzzling conundrums. Concerning the latter, baryon number violation appears fairly naturally in many new physics (NP) scenarios and is actually needed to explain the observed abundance of matter over antimatter in the Universe. Yet, the proton looks amazingly stable.

In this Letter, we observe that the flavor symmetries present in the SM gauge sector and their subsequent breaking through SM-like flavor mixings only, allow to lower the scale of the $\niente{B}$ and/or $\niente{L}$ violating (\blv) new physics at the TeV while still satisfying low-energy constraints like proton stability. Such \blv\ interactions could therefore be resonant at the energies probed by the LHC. To characterize their signatures in a model-independent way, effective field theory techniques can thus not be employed. We find, however, that  gauge and Lorentz invariant processes satisfying the three-generation flavor requirement can only involve very specific combinations of at least six external SM fermions, possibly accompanied by SM bosons or fermion-antifermion pairs. We denote by non-local fermionic channel a class of such six-fermion-based processes and single out two particular ones giving rise to striking signatures at the LHC: a single top in association with a muon-electron same-sign pair and two same-sign tops in association with jets, with a significant predominance of $\bar t\, \bar t$ over  $t\, t$. We consider realizations of such non-local fermionic channels in simplified models of new TeV-scale resonances that can satisfy low-energy and collider constraints and yet possibly lead to event rates accessible at the LHC.
\begin{figure*}[htb]
\centering
\includegraphics[width=\textwidth]{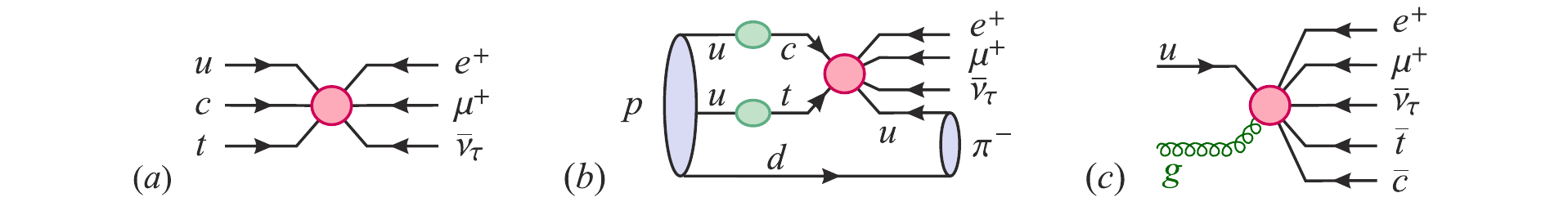}
\caption{Examples of six-fermion $B$ and/or $L$ violating processes. 
$(a)$~Three-generation fermionic core interaction. 
$(b)$~Proton decay process, where the suppression induced by the SM-like flavor transitions ensures a lifetime compatible with current limits. 
$(c)$~Unsuppressed flavor-diagonal $B$ and/or $L$ violating process accessible at the LHC.}
\label{Fig1}
\end{figure*}
\section{Flavor-diagonal \texorpdfstring{${\niente{B}}$}{B} and \texorpdfstring{${\niente{L}}$}{L} violation}
Our starting point is the flavor symmetry of the SM: the gauge interactions of the five fermion species (two left-handed fermion doublets, and three right-handed fermions per generation) are independent of their flavors. In other words, the kinetic Lagrangian of the SM exhibits a large $U(N_g)^5$ flavor symmetry~\cite{Chivukula:1987py}, with $N_g$ the number of generations. Crucially, even though this symmetry ends up broken in the SM, it happens in a very peculiar way that is extremely well supported experimentally. First, the explicit breaking  induced by the Yukawa interactions only, leads to highly hierarchical fermion masses and quark flavor mixings. As a result, SM flavor-changing transitions are mostly diagonal. Second, the $U(1)$ factors corresponding to the conventional $\niente{B}$ and $\niente{L}$ charges are accidentally conserved at the classical level and turn out to be broken by the SM dynamics through quantum effects. Yet these anomalous $\niente{B}-\niente{L}$ conserving interactions are extremely suppressed at low temperatures~\cite{'tHooft:1976up, *'tHooft:1976fv, *PhysRevD.18.2199.3}. So, even though NP scenarios may in principle introduce many new sources of breaking of the full flavor group, they cannot be generic: any new $SU(N_g)^5$ breaking term is required by flavor experiments to be roughly aligned with the SM ones and any new breaking of $U(1)_{\niente{B},\niente{L}}$ must pass the stringent bounds on the proton lifetime.

To introduce our strategy, let us first assume the exact conservation (\emph{flavor-blindness}) by the NP of the $SU(N_g)^5$ part of the full flavor symmetry group, but not its invariance under the $U(1)$ factors. It turns out that the $U(1)_{\niente{B}}$ and/or $U(1)_{\niente{L}}$ may be broken, but only in a very specific way. Indeed, the only $SU(N_g)^5$ flavor invariants are either identical fermion--antifermion pairs ($\delta_{a}^{b}\:\bar\psi^a\psi_b$) which conserve $\niente{B}$ and $\niente{L}$, or the combinations of $N_g$ identical fermions, antisymmetrized over their flavor indexes (e.g.,  $\epsilon^{abc}\psi_a\psi_b\psi_c$ when $N_g= 3$). Thus, flavor-blind \blv\ interactions are possible, yet only if they respect the selection rules: $N_c\,\Delta \niente{B} \in {N_g}\,\mathbb{Z}$ together with $\Delta \niente{L} \in N_g\, \mathbb{Z}$, where $N_c$ is the number of QCD colors and $\mathbb{Z}$ the set of integers~\cite{Smith:2011rp}. Remarkably, with $N_g=N_c$, $\Delta \niente{B}$ can take any integer value. Assuming $N_g=N_c=3$ and imposing the SM gauge and Lorentz invariances imply that the simplest flavor-blind \blv\ processes involve at least twelve fermions~\cite{Smith:2011rp}, i.e., as many as for the anomalous $B-L$ conserving interaction of the SM~\cite{'tHooft:1976up, *'tHooft:1976fv, *PhysRevD.18.2199.3}.

A strict flavor-blind setting is, however, not realistic as the $SU(3)^5$ flavor symmetry is already broken in the SM by the fermion mass spectrum. Interestingly, if the NP dynamics does not introduce new significant sources of flavor symmetry breaking, which is well supported experimentally, then the selection rules given above are unaffected, but six-fermion interactions (see Fig.~\ref{Fig1}$(a)$) become allowed~\cite{Smith:2011rp}. As a result, the simplest possible \blv\ interactions are characterized by $(\Delta \niente{B}; \Delta \niente{L}) = (0;\pm 6),$ $(\pm 1;\pm 3),$ $(\pm 1;\mp 3),$ $(\pm 2; 0)$.

\subsection{Low-energy constraints}

The absence of new significant $SU(3)^5$ symmetry breaking sources in the NP dynamics ensures not only that any TeV-scale NP scenario passes the constraints from flavor factories, but also that the expected proton lifetime remains above the current experimental limits. First, it automatically forbids the four-fermion operators~\cite{Weinberg:1979sa, *Wilczek:1979hc, *Abbott:1980zj} which induce $(\Delta\niente{B};\Delta\niente{L})=(\pm1;\pm1)$ transitions that are very much constrained at low energies. Second, low-energy constraints on six-fermion effective operators are satisfied thanks to their high dimensionality (leading to a suppression factor $(m_\text{nucleon}/1\: \text{TeV})^{10} \approx 10^{-30}$ for decay rates), and to the presence of antisymmetric contractions in flavor space. At first sight, these contractions require all the flavors, including the heaviest, to be simultaneously present, so that (di)nucleon decays or neutron--antineutron oscillations would end up kinematically forbidden. In fact, the presence of the SM flavor breaking terms allows for mixings so that \blv\ processes among light quarks and leptons are not strictly forbidden (see Fig.~\ref{Fig1}$(b)$). However, as long as flavor mixings remain SM-like, the flavor suppression is such that all the experimental constraints are satisfied, even with NP at the TeV scale~\cite{Smith:2011rp}.

\begin{table*}
\newcommand{\ms}{\quad}
\newcommand{\p}{$+$} %{$>0$}
\newcommand{\m}{$-$} %{$<0$}
\newlength{\intercol}
\settowidth{\intercol}{\qquad\qquad}
\newcommand{\ic}{\hspace{\intercol}}
\begin{center}
\begin{tabular}[c]{%
c@{\;}c @{\ic}
%	r@{\,$\otimes$\,}l @{\ic}
	r@{\,\,}l @{\ic}
		r@{\,$\otimes$\,}l @{\ic}
			l @{\ic}
				c%@{\ic}
					%c@{\ic}c
					}\hline\hline
$\Delta B$&$\Delta L$ 
	& \multicolumn{2}{c}{\hspace{-\intercol}Fermionic cores} 
		& \multicolumn{2}{c}{\hspace{-\intercol}Examples}
			& Promising LHC processes 
				& $A_{e\mu}$
					\\\hline
$0$&$\pm6$ 
	& NNN&NNN 
		& $\nu_{e}\,\nu_{\mu}\,\nu_{\tau}$&$\nu_{e}\,\nu_{\mu}\,\nu_{\tau}$
			& $u\,\bar{u}\to e^{-}\mu^{-}\nu_{\tau}\nu_{e}\nu_{\mu}\nu_{\tau}\ms W^{+}W^{+}\;$ 
				& $0$
				\\\hline
$\pm1$&$\pm3$ 
	& UUU&EEN
		& $t\,c\,u$&$e^{-}\,\mu^{-}\,\nu_{\tau}$
			& $u\,c\to\bar{t}\ms e^{+}\,\mu^{+}\,\bar{\nu}_{\tau}$ 
				& \p
				\\
&	&\omit&	&\omit&	& $u\,g\to\bar{t}\,\bar{c}\ms e^{+}\,\mu^{+}\,\bar{\nu}_{\tau}$
				& \p
				\\
&	&\omit&	&\omit&	& $g\,g\to\bar{t}\,\bar{c}\,\bar{u}\ms e^{+}\,\mu^{+}\,\bar{\nu}_{\tau}$
				& $0$
				\\
&	&\omit&	&\omit&	& $u\,c\to\bar{t}\ms e^{+}\,\mu^{+}\,\tau^{+}\ms W^{-}$
				& \p
				\\
&	& UUD&ENN
		& $t\,c\,d$&$e^{-}\,\nu_{\mu}\,\nu_{\tau}$
			& $d\,c\to\bar{t}\ms e^{+}\,\mu^{+}\,\bar \nu_{\tau}\ms W^{-}$
				& \p
				\\
&	& UDD&NNN 
		& $t\,s\,d$&$\nu_{e}\,\nu_{\mu}\,\nu_{\tau}$
			& $d\,s\to\bar{t}\ms e^{+}\,\mu^{+}\,\bar \nu_{\tau}\ms W^{-}W^{-}$
				& \p
				\\\hline
$\pm1$&$\mp3$
	& UDD&\=N\=N\=N
		& $t\,s\,d$&$\bar\nu_{e}\,\bar\nu_{\mu}\,\bar\nu_\tau$
			& $d\,s\to\bar t \ms e^-\,\mu^-\,\nu_\tau\ms W^{+}W^{+}$
				& \m
				\\
&	& DDD&\=E\=N\=N
		& $b\,s\,d$&$e^{+}\,\bar\nu_\mu\,\bar\nu_\tau$
			& $d\,s\to\bar t \ms e^-\,\mu^-\,\nu_{\tau}\ms W^{+}W^{+}$
				& \m
				\\\hline
$\pm2$&$0$
	& UDD&UDD
		& $t\,s\,d$&$t\,s\,d$
			& $d\,d\to\bar{t}\,\bar{t}\ms \bar{s}\,\bar{s}$
				& \m
				\\
&	&\omit&	&\omit&	& $d\,g\to\bar{t}\,\bar{t}\ms \bar{s}\,\bar{s}\ms \bar{d}$ 
				& \m
				\\
&	&\omit&	&\omit&	& $g\,g\to\bar{t}\,\bar{t}\ms \bar{s}\,\bar{s}\ms \bar{d}\,\bar{d}$
				& $0$
				\\
&	& &%UUD&DDD
		& $t\,c\,d$&$b\,s\,d$
			& $d\,u\rightarrow\bar{t}\,\bar{t}\ms \bar{s}\,\bar{s}\ms W^{+}$
				& \m
				\\
&	&\omit&	&\omit&	& $d\,d\to\bar{t}\,\bar{t}\ms \bar{c}\,\bar{s}\ms W^{+}$
				& \m
				\\\hline\hline
\end{tabular}
\end{center}
\caption{\emph{First two columns}: The selection rules for flavor-diagonal $B$ and/or $L$ violation and corresponding six-fermion cores. U, D, E, and N are flavor-generic up-, down-type quarks, charged lepton and neutrino, respectively. %No distinction is made between left and right-handed fields. 
Charge-conjugate interactions are understood as well as antisymmetrization over the quark or lepton flavor indexes.
%For example, UUD$\,\otimes\,$ENN stands for $\epsilon^{abc}\text{U}_a \text{U}_b \text{D}_c \; \epsilon^{def}\text{E}_d \text{N}_e \text{N}_f$. %, where flavor indexes $a, b,... = 1,2,3$ are summed over.
\emph{Third column}: Example of non-local fermionic channels corresponding to each fermionic core, with specific flavor assignments.
\emph{Fourth column}: Promising  flavor-diagonal $B$ and/or $L$ violating transitions to look for at the LHC for each non-local fermionic channel.
\emph{Fifth column}: The expected dilepton charge asymmetry, as defined in Eq.~(\ref{Eq3}), for an $e\mu$ final-state pair.
%\emph{Last column}: Cross-sections predicted in the simplified models discussed.
}
\label{tab-fec}
\label{tab-proc}
\end{table*}

\subsection{Non-local fermionic channels}

At TeV colliders, there is no need to call in for flavor mixing suppressions since heavy fermions can be produced directly. So, \emph{flavor-diagonal} \blv\ transitions, i.e., transitions involving all three generations and surviving in the limit where all Yukawa couplings are simultaneously diagonal, are readily accessible (see Fig.~\ref{Fig1}$(c)$). In other words, \blv\ interactions are flavor-diagonal when explicitly antisymmetric in flavor space. Furthermore, since low-energy constraints are compatible with new dynamics at the TeV scale, such processes could be resonant (i.e., non-local) at colliders. Were they not resonant, no signal would actually be visible, essentially because local interactions involving at least six fermions are extremely suppressed by powers of the NP scale and by phase space.

We therefore assume that NP is at the TeV scale and that \blv\ transitions can be mediated by resonant states promptly decaying into SM fields. Restricting ourselves first to the minimal content of six SM fermions that would actually be seen in detectors once NP states have decayed, we only get a handful of field combinations allowed by the overall color and electric charge conservation as well as by flavor-diagonality (see the second column of Table~\ref{tab-fec}). In a resonant regime, the underlying NP dynamics can no longer be approximated as local and each of these minimal fermionic cores can actually be considered as the representative of a larger class of possible processes we name a \emph{non-local fermionic channel} (\nlfc). They are not effective operators. Specifically, because the power counting of effective field theories can no longer be used, each of these \nlfc\ may include additional flavor-diagonal, $\niente{B}$ and $\niente{L}$ conserving, combinations of SM particles beside the six core fermions. Thus, they represent the channels with six fermions alone or together with neutral SM gauge bosons, fermion-antifermion pairs, neutral Higgs bosons, or with one or several of the core fermions replaced by their $SU(2)_L$ partners and a $W$, for instance $\text{N}\to\text{E}W^+$ or $\text{D}\to\text{U}W^-$ (see the fourth column of Table~\ref{tab-fec}).

Notably, the fermionic cores of Table~\ref{tab-fec} all involve \emph{same-sign} combinations of leptons and quarks, i.e., either only leptons or only antileptons, and, either only quarks or only antiquarks. This characteristic feature is preserved at the \nlfc\ level, in the presence of additional external SM bosons. At colliders, the lepton and top-quark charges therefore provide a crucial handle to disentangle \blv\ processes from SM background and other NP scenarios.
\section{LHC signatures}
We now focus on the LHC signals induced by \blv\ transitions~\cite{Dong:2011rh,*Baldes:2011mh} in our three-generation context. Despite the fact that each \nlfc\ actually involves more processes than the simplest six-fermion interactions, a well-defined search strategy can be established. To this end, we identify two main observables that rely on the same-sign character of final-state leptons and initial-state quarks, respectively.
 
\subsection{Same-sign leptons}

To detect three-generation \blv\ interactions at the LHC, the final state should satisfy some experimentally-driven criteria. First, having access to the charge of produced particles is of critical importance. A signature featuring a same-sign top pair or a same-sign electron-muon pair at the LHC for instance fulfills this criterion. In addition, it also permits to avoid large SM backgrounds. Second, signatures with a minimal number of light quark jets and neutrinos are preferable.

These phenomenological considerations strongly constrain the number of viable channels (see the fourth column of Table~\ref{tab-proc}). On the one hand, the $(\Delta\niente{B};\Delta\niente{L})=(0;\pm6)$ and $(\pm 1;\mp3)$ processes are difficult to reach. Neutrinos could be replaced by $\ell^{-}W^{+}$ fields, with the $W$ possibly decaying hadronically not to change the characteristic leptonic signature. Even though it is in principle possible that a \blv\ transition occurs only concurrently with, or resonate only in the presence of, the emission of some SM bosons, this might turn out to be experimentally more difficult to identify. On the other hand, the $\text{UUU}\,\,\text{EEN}$ and $\text{UDD}\,\,\text{UDD}$ fermionic cores make the $(\Delta\niente{B};\Delta\niente{L})=(\pm1;\pm3)$ and $(\pm 2;0)$ transitions readily accessible at the LHC through the processes displayed in Table~\ref{tab-proc}.

\subsection{Charge asymmetries}

As a proton-proton collider, the LHC gives the possibility to build non-trivial observables based on the electric charge information. Valence quarks appear more often than sea quarks in the initial state. Therefore, $qq$ initial states dominate over $\bar q\bar q$ ones and this asymmetry may propagate in the final state. In practice, this effect driven by parton distribution functions (PDFs) is characteristic of all processes of Table~\ref{tab-proc} and can be quantified by means of a dilepton charge asymmetry, defined as
\begin{equation}
A_{\ell_a\ell_b}
\equiv
\frac{
	\sigma\np(p\:p\to\ell_{a}^{+}\ell_{b}^{+}X)-
	\sigma\np(p\:p\to\ell_{a}^{-}\ell_{b}^{-}\bar X)
}{
	\sigma\np(p\:p\to\ell_{a}^{+}\ell_{b}^{+}X)+
	\sigma\np(p\:p\to\ell_{a}^{-}\ell_{b}^{-}\bar X)
}\;,
\label{Eq3}
\end{equation}
which can be close to $+1$ or $-1$ for some lepton flavors (denoted by $a,b$). Thus, these charge asymmetries, especially when combined with the leptonic flavor information, provide characteristic signals for flavor-diagonal \blv\ interactions, and allow to discriminate among different scenarios (see the last column of Table~\ref{tab-proc}). 

\begin{figure*}
\includegraphics[width=\textwidth]{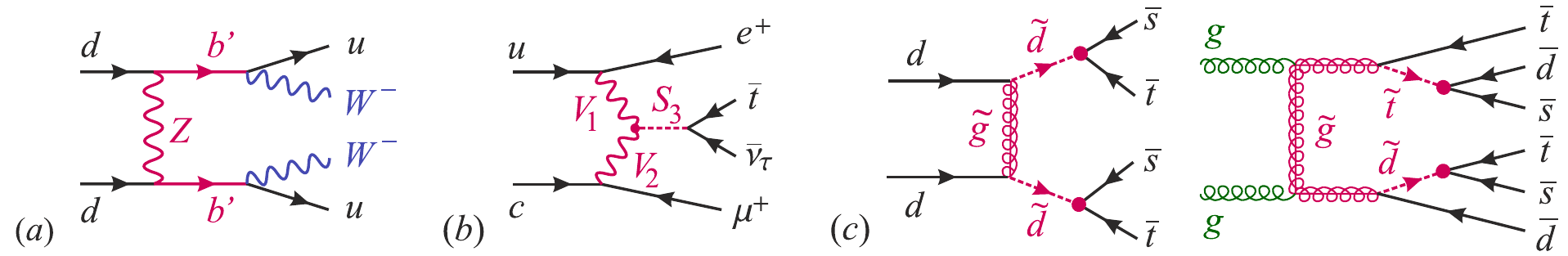}
\caption{$(a)$ $B$ and $L$ conserving process that might also give rise to a negative $A_{\ell_a\ell_b}$ charge asymmetry in NP production rates of same-sign prompt and isolated dilepton. It involves flavor-changing neutral currents and a new heavy down-type quark denoted by $b'$ that couples preferentially to the first generation.
$(b)$ Example of $(\Delta\niente{B};\Delta\niente{L})=(\pm1,\pm3)$ process with $A\LQ_{e\mu}>0$ in our leptoquark simplified model.
$(c)$ Examples of $R$-parity violating flavor-diagonal $(\Delta \niente{B}; \Delta \niente{L})=(\pm 2;0)$ processes with $A\RPV_{e\mu}\le 0$. Quark (gluon) initiated transitions are likely to dominate if squarks (gluinos) are light and therefore resonate.}
\label{fig-bprime}
\label{fig-lq}
\label{fig-rpv}
\end{figure*}

For example, in the $(\Delta\niente{B};\Delta\niente{L})=(\pm2;0)$ case, the $d\,d$ initial state of $p\,p\to \bar t\,\bar t\,\bar s\,\bar s + X$ is about $20$ ($100$) times~\cite{details} more likely than the $\bar d\,\bar d$ one of $p\,p\to t\,t\,s\,s + \bar X$, for a partonic center of mass energy of $1$ ($2$)~TeV, at the $8$~TeV LHC, with CTEQ6L1 PDFs~\cite{Pumplin:2002vw}. Thus, looking at the leptonic decays of the two tops, such \blv\ processes will lead to many more negatively charged same-sign isolated leptons than positively charged ones. Since decaying top quarks produce all lepton flavors in equal proportions, $A_{\ell_a\ell_b}$ would be close to $-1$ for all $a, b$. It should be stressed that this is a very characteristic signal. Same-sign isolated dileptons feature a positive charge asymmetry in the SM as well as in most NP models. Other NP scenarios leading to a negative one are quite peculiar and seem to require flavor-changing neutral currents in the down sector (see Fig.~\ref{fig-bprime}$(a)$). 

The $(\Delta\niente{B};\Delta\niente{L})=(\pm1;\pm3)$ and $(\pm1;\mp 3)$ interactions also lead to charge asymmetries, but only for leptons of different flavors, i.e., $A_{\ell_a\ell_b}$ could be close to $\pm1$ for $a\neq b$. These leptons are directly produced by the NP dynamics and are therefore required to be in a flavor-diagonal combination. Even the processes characterized by $(\Delta\niente{B};\Delta\niente{L})=(\pm1;\pm3)$, for which $A_{\ell_a\ell_b}$ is positive, are clearly identifiable as NP.

Finally, it should be mentioned that \blv\ interactions do not necessarily generate an asymmetry. Gluons can replace light quarks in the initial state. If not too severely suppressed by phase space, such processes could be PDF-enhanced, or may resonate more easily than the quark-initiated ones. In that case, the neutral initial state would render the charge asymmetry vanishing. Still, the value of the asymmetry would carry important information about the NP dynamics as illustrated now.
\subsection{Simplified \texorpdfstring{${\niente{B}}$}{B} and \texorpdfstring{${\niente{L}}$}{L} violating models for the LHC}
To get estimates of flavor-diagonal \blv\ rates, we consider simplified models giving rise to the two particularly interesting \nlfc\ characterized by $(\Delta\niente{B};\Delta\niente{L})=(\pm1;\pm3)$ and $(\pm2;0)$, with $A_{e\mu}\ge 0$ and $A_{e\mu}\le 0$ respectively, $A_{e\mu}<0$ being a smoking gun.

\paragraph{Leptoquark simplified model}

Both vector and scalar leptoquarks are required to give rise to the $t\,c\,u\,\otimes\,e^-\,\mu^-\,\nu_\tau$ \nlfc. As our purpose is illustrative only, we do not address the construction of a fully dynamical leptoquark model but restrict ourselves to a simplified one (see Fig.~\ref{fig-lq}$(b)$). Vector (scalar) leptoquarks are taken to couple chirally to a charged lepton (neutrino) and an up-type quark. Couplings to fermions of the same generation are assumed dominant and their strength is fixed universally to $0.6$. For given vector leptoquark masses this coupling constant is constrained by the $e^+e^-$ and $\mu^+\mu^-$ observed spectra~\cite{CMS:2012exa,*ATLAS:2012ipa}. A trilinear vector--vector--scalar vertex between leptoquarks has a completely antisymmetric flavor structure and gives rise to the three-generation signature. Vectors, with Yang-Mills couplings to gluons, have a common mass of $1$~TeV while the scalars masses are taken to be $500$~GeV. Existing QCD pair production searches~\cite{:2012dnb,*CMS-PAS-EXO-11-030,*CMS-PAS-EXO-12-002,*Aad:2011ch,*ATLAS:2012aq} do not target explicitly the leptoquarks considered here but we have chosen values for masses possibly compatible with an absence of signal in the available data. At the $8$ ($14$)~TeV LHC, total rates estimated with the \textsc{FeynRules}-\textsc{MadGraph5}~\cite{Christensen:2008py,Alwall:2011uj} software chain are found to be~\cite{details}:
\begin{equation*}
%GEN_LQ_uc_emvt_5
%	& .0028759000\:	(.0248320000)
%	& .9337681549\:	(.9025946910)
%GEN_LQ_ug_emvtc_5
%	& .0176663000\:	(.40286741000)
%	& .9560109834\:	(.9425135347)
%GEN_LQ_gg_emvtcu_5
%	& .00193746000\:	(.252535723769000)
%	& .0009971429\:	(-.0009948085)
\begin{array}{rl}
\sigma\LQ(u\,c\to\bar t\;\; e^+\,\mu^+\bar\nu_\tau) =
	& 0.0029\: (0.025)\fb,
\\[-.5mm]
A\LQ_{e\mu} = & +0.93\: (+0.90)
\\[1mm]
\sigma\LQ(u\,g\to\bar t\,\bar c\;\; e^+\,\mu^+\,\bar\nu_\tau) =
	& 0.018\: (0.40)\fb,
\\[-.5mm]
A\LQ_{e\mu} = & +0.96\: (+0.94)
\\[1mm]
\sigma\LQ(g\,g\to\bar t\,\bar c\,\bar u \;\; e^+\,\mu^+\,\bar\nu_\tau) =
	& 0.0019\: (0.25)\fb.
\end{array}
\end{equation*}
\paragraph{R-parity violating simplified model\label{rpv}}

We consider a simplified supersymmetric extension of the SM restricted to the super-QCD sector with all squarks degenerate and one single sizable R-parity violating (RPV) coupling: $\lambda''_{tds}=0.1$~\cite{Nikolidakis:2007fc}. Rates depend on the mass hierarchy between gluino and squarks (see Fig.~\ref{fig-rpv}$(c)$). We choose two benchmark hierarchies for which the most relevant low energy and collider constraints are satisfied. Despite instrumental (charge, jet, heavy meson, or photon conversion misidentification) as well as irreducible ($t\,\bar t\,W$, $t\,\bar t\, Z$, di- or tri-bosons) backgrounds, existing same-sign isolated dilepton searches at the LHC~\cite{:2012vh,*ATLAS:2012sna} turn out to be fairly constraining~\cite{details}. As in the leptoquark case, we estimate the $8$ ($14$)~TeV LHC charge asymmetries and total rates~\cite{details}:
\begin{equation*}
\begin{array}{rcc}
\begin{array}{c}
\sigma\RPV\:\:
\end{array}
	& \begin{array}{c}
	  m_{\tilde q} = 600,\\[-1mm]
	  m_{\tilde g} = 750,
	  \end{array}
	  	& \begin{array}{c}
	  	  800\gev,\\[-1mm]
	  	  650\gev,
	  	  \end{array} \\[3mm]
d\,d\to\bar t\,\bar t\:%+
\bar s\bar s
\,:
	& 30\:(110),
		& 0.012\: (0.065)\fb,\\[-.5mm]
A_{\ell_a\ell_b}\RPV =
	& -0.95\:(-0.88),
		& -0.98\:(-0.92),\\[1mm]
g\,d\to\bar t\,\bar t\:%+
\bar s\bar s\:
\bar d
\,:
	& 16\:(140),
		& 1.2\: (12)\fb,\\[-.5mm]
A_{\ell_a\ell_b}\RPV =
	& -0.80\:(-0.69),
		& -0.81\:(-0.69),\\[1mm]
g\,g\to\bar t\,\bar t\:%+
\bar s\,\bar s\:%+
\bar d\, \bar d
\,:
	& 1.7\:(33),
		& 38\: (590)\fb.
\end{array}
\end{equation*}
\section{Conclusions}
We have argued that, if compatible with Standard Model flavor symmetries, baryon and/or lepton number violation can arise at scales as low as the TeV without conflicting with stringent low-energy constraints. At colliders able to produce heavy fermions directly, such a violation would involve simultaneously all three quark and/or lepton generations. We have classified corresponding LHC signals in a fully model-independent way and found that same-sign dileptons, their flavor, and the charge asymmetry in their production rate provide unique handles to discriminate between $B$ and/or $L$ violating processes and SM backgrounds or other new physics scenarios. This has been illustrated with two simplified models: a generic leptoquark setting and a restricted R-parity violating supersymmetric model. Remarkably, both of them are already forced by existing LHC searches to arise at scales no lower than a fraction of TeV. With dedicated studies and more data, the LHC will therefore offer us a fantastic opportunity to finally unravel the true status of the $B$ and $L$ quantum numbers in Nature.

\section*{Acknowledgments}
This research has been supported in part by the Belgian IAP
Program BELSPO P7/37. G.D. is a Research Fellow of the F.R.S.--FNRS, Belgium.

\section*{References}
\bibliographystyle{apsrev4-1}
\bibliography{refs}

%merlin.mbs apsrev4-1.bst 2010-07-25 4.21a (PWD, AO, DPC) hacked
%Control: key (0)
%Control: author (72) initials jnrlst
%Control: editor formatted (1) identically to author
%Control: production of article title (-1) disabled
%Control: page (0) single
%Control: year (1) truncated
%Control: production of eprint (0) enabled
\begin{thebibliography}{24}%
\makeatletter
\providecommand \@ifxundefined [1]{%
 \@ifx{#1\undefined}
}%
\providecommand \@ifnum [1]{%
 \ifnum #1\expandafter \@firstoftwo
 \else \expandafter \@secondoftwo
 \fi
}%
\providecommand \@ifx [1]{%
 \ifx #1\expandafter \@firstoftwo
 \else \expandafter \@secondoftwo
 \fi
}%
\providecommand \natexlab [1]{#1}%
\providecommand \enquote  [1]{``#1''}%
\providecommand \bibnamefont  [1]{#1}%
\providecommand \bibfnamefont [1]{#1}%
\providecommand \citenamefont [1]{#1}%
\providecommand \href@noop  [0]{\@secondoftwo}%
\providecommand \href [0]{\begingroup \@sanitize@url \@href}%
\providecommand \@href[1]{\@@startlink{#1}\@@href}%
\providecommand \@@href[1]{\endgroup#1\@@endlink}%
\providecommand \@sanitize@url [0]{\catcode `\\12\catcode `\$12\catcode
  `\&12\catcode `\#12\catcode `\^12\catcode `\_12\catcode `\%12\relax}%
\providecommand \@@startlink[1]{}%
\providecommand \@@endlink[0]{}%
\providecommand \url  [0]{\begingroup\@sanitize@url \@url }%
\providecommand \@url [1]{\endgroup\@href {#1}{\urlprefix }}%
\providecommand \urlprefix  [0]{URL }%
\providecommand \href [0]{\href }%
\providecommand \doibase [0]{http://dx.doi.org/}%
\providecommand \selectlanguage [0]{\@gobble}%
\providecommand \bibinfo  [0]{\@secondoftwo}%
\providecommand \bibfield  [0]{\@secondoftwo}%
\providecommand \translation [1]{[#1]}%
\providecommand \BibitemOpen [0]{}%
\providecommand \bibitemStop [0]{}%
\providecommand \bibitemNoStop [0]{.\EOS\space}%
\providecommand \EOS [0]{\spacefactor3000\relax}%
\providecommand \BibitemShut  [1]{\csname bibitem#1\endcsname}%
\let\auto@bib@innerbib\@empty
%</preamble>
\bibitem [{\citenamefont {Chivukula}\ and\ \citenamefont
  {Georgi}(1987)}]{Chivukula:1987py}%
  \BibitemOpen
  \bibfield  {author} {\bibinfo {author} {\bibfnamefont {R.~S.}\ \bibnamefont
  {Chivukula}}\ and\ \bibinfo {author} {\bibfnamefont {H.}~\bibnamefont
  {Georgi}},\ }\href {\doibase 10.1016/0370-2693(87)90713-1} {\bibfield
  {journal} {\bibinfo  {journal} {Phys.Lett.}\ }\textbf {\bibinfo {volume}
  {B188}},\ \bibinfo {pages} {99} (\bibinfo {year} {1987})}\BibitemShut
  {NoStop}%
%%CITATION = PHLTA,B188,99;%%
\bibitem [{\citenamefont {'t~Hooft}(1976{\natexlab{a}})}]{'tHooft:1976up}%
  \BibitemOpen
  \bibfield  {author} {\bibinfo {author} {\bibfnamefont {G.}~\bibnamefont
  {'t~Hooft}},\ }\href {\doibase 10.1103/PhysRevLett.37.8} {\bibfield
  {journal} {\bibinfo  {journal} {Phys.Rev.Lett.}\ }\textbf {\bibinfo {volume}
  {37}},\ \bibinfo {pages} {8} (\bibinfo {year}
  {1976}{\natexlab{a}})}\BibitemShut {NoStop}%
%%CITATION = PRLTA,37,8;%%
\bibitem [{\citenamefont {'t~Hooft}(1976{\natexlab{b}})}]{'tHooft:1976fv}%
  \BibitemOpen
  \bibfield  {author} {\bibinfo {author} {\bibfnamefont {G.}~\bibnamefont
  {'t~Hooft}},\ }\href {\doibase 10.1103/PhysRevD.14.3432} {\bibfield
  {journal} {\bibinfo  {journal} {Phys.Rev.}\ }\textbf {\bibinfo {volume}
  {D14}},\ \bibinfo {pages} {3432} (\bibinfo {year}
  {1976}{\natexlab{b}})}\BibitemShut {NoStop}%
%%CITATION = PHRVA,D14,3432;%%
\bibitem [{\citenamefont {'t~Hooft}(1978)}]{PhysRevD.18.2199.3}%
  \BibitemOpen
  \bibfield  {author} {\bibinfo {author} {\bibfnamefont {G.}~\bibnamefont
  {'t~Hooft}},\ }\href {\doibase 10.1103/PhysRevD.18.2199.3} {\bibfield
  {journal} {\bibinfo  {journal} {Phys.Rev.}\ }\textbf {\bibinfo {volume}
  {D18}},\ \bibinfo {pages} {2199} (\bibinfo {year} {1978})}\BibitemShut
  {NoStop}%
\bibitem [{\citenamefont {Smith}(2012)}]{Smith:2011rp}%
  \BibitemOpen
  \bibfield  {author} {\bibinfo {author} {\bibfnamefont {C.}~\bibnamefont
  {Smith}},\ }\href {\doibase 10.1103/PhysRevD.85.036005} {\bibfield  {journal}
  {\bibinfo  {journal} {Phys.Rev.}\ }\textbf {\bibinfo {volume} {D85}},\
  \bibinfo {pages} {036005} (\bibinfo {year} {2012})},\ \href
  {http://arxiv.org/abs/1105.1723}{arXiv:1105.1723 [hep-ph]}\BibitemShut
  {NoStop}%
%%CITATION = ARXIV:1105.1723;%%
\bibitem [{\citenamefont {Weinberg}(1979)}]{Weinberg:1979sa}%
  \BibitemOpen
  \bibfield  {author} {\bibinfo {author} {\bibfnamefont {S.}~\bibnamefont
  {Weinberg}},\ }\href {\doibase 10.1103/PhysRevLett.43.1566} {\bibfield
  {journal} {\bibinfo  {journal} {Phys.Rev.Lett.}\ }\textbf {\bibinfo {volume}
  {43}},\ \bibinfo {pages} {1566} (\bibinfo {year} {1979})}\BibitemShut
  {NoStop}%
%%CITATION = PRLTA,43,1566;%%
\bibitem [{\citenamefont {Wilczek}\ and\ \citenamefont
  {Zee}(1979)}]{Wilczek:1979hc}%
  \BibitemOpen
  \bibfield  {author} {\bibinfo {author} {\bibfnamefont {F.}~\bibnamefont
  {Wilczek}}\ and\ \bibinfo {author} {\bibfnamefont {A.}~\bibnamefont {Zee}},\
  }\href {\doibase 10.1103/PhysRevLett.43.1571} {\bibfield  {journal} {\bibinfo
   {journal} {Phys.Rev.Lett.}\ }\textbf {\bibinfo {volume} {43}},\ \bibinfo
  {pages} {1571} (\bibinfo {year} {1979})}\BibitemShut {NoStop}%
%%CITATION = PRLTA,43,1571;%%
\bibitem [{\citenamefont {Abbott}\ and\ \citenamefont
  {Wise}(1980)}]{Abbott:1980zj}%
  \BibitemOpen
  \bibfield  {author} {\bibinfo {author} {\bibfnamefont {L.}~\bibnamefont
  {Abbott}}\ and\ \bibinfo {author} {\bibfnamefont {M.~B.}\ \bibnamefont
  {Wise}},\ }\href {\doibase 10.1103/PhysRevD.22.2208} {\bibfield  {journal}
  {\bibinfo  {journal} {Phys.Rev.}\ }\textbf {\bibinfo {volume} {D22}},\
  \bibinfo {pages} {2208} (\bibinfo {year} {1980})}\BibitemShut {NoStop}%
%%CITATION = PHRVA,D22,2208;%%
\bibitem [{\citenamefont {Dong}\ \emph {et~al.}(2012)\citenamefont {Dong},
  \citenamefont {Durieux}, \citenamefont {Gérard}, \citenamefont {Han},\ and\
  \citenamefont {Maltoni}}]{Dong:2011rh}%
  \BibitemOpen
  \bibfield  {author} {\bibinfo {author} {\bibfnamefont {Z.}~\bibnamefont
  {Dong}}, \bibinfo {author} {\bibfnamefont {G.}~\bibnamefont {Durieux}},
  \bibinfo {author} {\bibfnamefont {J.-M.}\ \bibnamefont {Gérard}}, \bibinfo
  {author} {\bibfnamefont {T.}~\bibnamefont {Han}}, \ and\ \bibinfo {author}
  {\bibfnamefont {F.}~\bibnamefont {Maltoni}},\ }\href {\doibase
  10.1103/PhysRevD.85.016006} {\bibfield  {journal} {\bibinfo  {journal}
  {Phys.Rev.}\ }\textbf {\bibinfo {volume} {D85}},\ \bibinfo {pages} {016006}
  (\bibinfo {year} {2012})},\ \href
  {http://arxiv.org/abs/1107.3805}{arXiv:1107.3805 [hep-ph]}\BibitemShut
  {NoStop}%
%%CITATION = ARXIV:1107.3805;%%
\bibitem [{\citenamefont {Baldes}\ \emph {et~al.}(2011)\citenamefont {Baldes},
  \citenamefont {Bell},\ and\ \citenamefont {Volkas}}]{Baldes:2011mh}%
  \BibitemOpen
  \bibfield  {author} {\bibinfo {author} {\bibfnamefont {I.}~\bibnamefont
  {Baldes}}, \bibinfo {author} {\bibfnamefont {N.~F.}\ \bibnamefont {Bell}}, \
  and\ \bibinfo {author} {\bibfnamefont {R.~R.}\ \bibnamefont {Volkas}},\
  }\href {\doibase 10.1103/PhysRevD.84.115019} {\bibfield  {journal} {\bibinfo
  {journal} {Phys.Rev.}\ }\textbf {\bibinfo {volume} {D84}},\ \bibinfo {pages}
  {115019} (\bibinfo {year} {2011})},\ \href
  {http://arxiv.org/abs/1110.4450}{arXiv:1110.4450 [hep-ph]}\BibitemShut
  {NoStop}%
%%CITATION = ARXIV:1110.4450;%%
\bibitem [{\citenamefont {Durieux}\ \emph {et~al.}()\citenamefont {Durieux},
  \citenamefont {Gérard}, \citenamefont {Maltoni},\ and\ \citenamefont
  {Smith}}]{details}%
  \BibitemOpen
  \bibfield  {author} {\bibinfo {author} {\bibfnamefont {G.}~\bibnamefont
  {Durieux}}, \bibinfo {author} {\bibfnamefont {J.-M.}\ \bibnamefont
  {Gérard}}, \bibinfo {author} {\bibfnamefont {F.}~\bibnamefont {Maltoni}}, \
  and\ \bibinfo {author} {\bibfnamefont {C.}~\bibnamefont {Smith}},\
  }\href@noop {} {\ }\bibinfo {note} {In preparation}\BibitemShut {NoStop}%
\bibitem [{\citenamefont {Pumplin}\ \emph {et~al.}(2002)\citenamefont
  {Pumplin}, \citenamefont {Stump}, \citenamefont {Huston}, \citenamefont
  {Lai}, \citenamefont {Nadolsky} \emph {et~al.}}]{Pumplin:2002vw}%
  \BibitemOpen
  \bibfield  {author} {\bibinfo {author} {\bibfnamefont {J.}~\bibnamefont
  {Pumplin}}, \bibinfo {author} {\bibfnamefont {D.}~\bibnamefont {Stump}},
  \bibinfo {author} {\bibfnamefont {J.}~\bibnamefont {Huston}}, \bibinfo
  {author} {\bibfnamefont {H.}~\bibnamefont {Lai}}, \bibinfo {author}
  {\bibfnamefont {P.~M.}\ \bibnamefont {Nadolsky}},  \emph {et~al.},\ }\href
  {\doibase 10.1088/1126-6708/2002/07/012} {\bibfield  {journal} {\bibinfo
  {journal} {JHEP}\ }\textbf {\bibinfo {volume} {0207}},\ \bibinfo {pages}
  {012} (\bibinfo {year} {2002})},\ \href
  {http://arxiv.org/abs/hep-ph/0201195}{arXiv:hep-ph/0201195
  [hep-ph]}\BibitemShut {NoStop}%
%%CITATION = HEP-PH/0201195;%%
\bibitem [{CMS(2012{\natexlab{a}})}]{CMS:2012exa}%
  \BibitemOpen
  \href {http://cds.cern.ch/record/1461216} {\bibfield  {journal} {\bibinfo
  {journal} {CMS-PAS-EXO-12-015}\ } (\bibinfo {year} {Jul
  2012}{\natexlab{a}})}\BibitemShut {NoStop}%
\bibitem [{ATL(2012{\natexlab{a}})}]{ATLAS:2012ipa}%
  \BibitemOpen
  \href {https://cds.cern.ch/record/1477926} {\bibfield  {journal} {\bibinfo
  {journal} {ATLAS-CONF-129-2012}\ } (\bibinfo {year} {Sep
  2012}{\natexlab{a}})}\BibitemShut {NoStop}%
\bibitem [{\citenamefont {Chatrchyan}\ \emph
  {et~al.}(2012{\natexlab{a}})\citenamefont {Chatrchyan} \emph
  {et~al.}}]{:2012dnb}%
  \BibitemOpen
  \bibfield  {author} {\bibinfo {author} {\bibfnamefont {S.}~\bibnamefont
  {Chatrchyan}} \emph {et~al.} (\bibinfo {collaboration} {CMS Collaboration}),\
  }\href {\doibase 10.1103/PhysRevD.86.052013} {\bibfield  {journal} {\bibinfo
  {journal} {Phys.Rev.}\ }\textbf {\bibinfo {volume} {D86}},\ \bibinfo {pages}
  {052013} (\bibinfo {year} {2012}{\natexlab{a}})},\ \href
  {http://arxiv.org/abs/1207.5406}{arXiv:1207.5406 [hep-ex]}\BibitemShut
  {NoStop}%
%%CITATION = ARXIV:1207.5406;%%
\bibitem [{CMS(2012{\natexlab{b}})}]{CMS-PAS-EXO-11-030}%
  \BibitemOpen
  \href {http://cds.cern.ch/record/1416079} {\bibfield  {journal} {\bibinfo
  {journal} {CMS-PAS-EXO-11-030}\ } (\bibinfo {year} {Jan
  2012}{\natexlab{b}})}\BibitemShut {NoStop}%
\bibitem [{CMS(2012{\natexlab{c}})}]{CMS-PAS-EXO-12-002}%
  \BibitemOpen
  \href {http://cds.cern.ch/record/1460842} {\bibfield  {journal} {\bibinfo
  {journal} {CMS-PAS-EXO-12-002}\ } (\bibinfo {year} {Jul
  2012}{\natexlab{c}})}\BibitemShut {NoStop}%
\bibitem [{\citenamefont {Aad}\ \emph {et~al.}(2012{\natexlab{a}})\citenamefont
  {Aad} \emph {et~al.}}]{Aad:2011ch}%
  \BibitemOpen
  \bibfield  {author} {\bibinfo {author} {\bibfnamefont {G.}~\bibnamefont
  {Aad}} \emph {et~al.} (\bibinfo {collaboration} {ATLAS Collaboration}),\
  }\href {\doibase 10.1016/j.physletb.2012.02.004} {\bibfield  {journal}
  {\bibinfo  {journal} {Phys.Lett.}\ }\textbf {\bibinfo {volume} {B709}},\
  \bibinfo {pages} {158} (\bibinfo {year} {2012}{\natexlab{a}})},\ \href
  {http://arxiv.org/abs/1112.4828}{arXiv:1112.4828 [hep-ex]}\BibitemShut
  {NoStop}%
%%CITATION = ARXIV:1112.4828;%%
\bibitem [{\citenamefont {Aad}\ \emph {et~al.}(2012{\natexlab{b}})\citenamefont
  {Aad} \emph {et~al.}}]{ATLAS:2012aq}%
  \BibitemOpen
  \bibfield  {author} {\bibinfo {author} {\bibfnamefont {G.}~\bibnamefont
  {Aad}} \emph {et~al.} (\bibinfo {collaboration} {ATLAS Collaboration}),\
  }\href {\doibase 10.1140/epjc/s10052-012-2151-6} {\bibfield  {journal}
  {\bibinfo  {journal} {Eur.Phys.J.}\ }\textbf {\bibinfo {volume} {C72}},\
  \bibinfo {pages} {2151} (\bibinfo {year} {2012}{\natexlab{b}})},\ \href
  {http://arxiv.org/abs/1203.3172}{arXiv:1203.3172 [hep-ex]}\BibitemShut
  {NoStop}%
%%CITATION = ARXIV:1203.3172;%%
\bibitem [{\citenamefont {Christensen}\ and\ \citenamefont
  {Duhr}(2009)}]{Christensen:2008py}%
  \BibitemOpen
  \bibfield  {author} {\bibinfo {author} {\bibfnamefont {N.~D.}\ \bibnamefont
  {Christensen}}\ and\ \bibinfo {author} {\bibfnamefont {C.}~\bibnamefont
  {Duhr}},\ }\href {\doibase 10.1016/j.cpc.2009.02.018} {\bibfield  {journal}
  {\bibinfo  {journal} {Comput.Phys.Commun.}\ }\textbf {\bibinfo {volume}
  {180}},\ \bibinfo {pages} {1614} (\bibinfo {year} {2009})},\ \href
  {http://arxiv.org/abs/0806.4194}{arXiv:0806.4194 [hep-ph]}\BibitemShut
  {NoStop}%
%%CITATION = ARXIV:0806.4194;%%
\bibitem [{\citenamefont {Alwall}\ \emph {et~al.}(2011)\citenamefont {Alwall},
  \citenamefont {Herquet}, \citenamefont {Maltoni}, \citenamefont {Mattelaer},\
  and\ \citenamefont {Stelzer}}]{Alwall:2011uj}%
  \BibitemOpen
  \bibfield  {author} {\bibinfo {author} {\bibfnamefont {J.}~\bibnamefont
  {Alwall}}, \bibinfo {author} {\bibfnamefont {M.}~\bibnamefont {Herquet}},
  \bibinfo {author} {\bibfnamefont {F.}~\bibnamefont {Maltoni}}, \bibinfo
  {author} {\bibfnamefont {O.}~\bibnamefont {Mattelaer}}, \ and\ \bibinfo
  {author} {\bibfnamefont {T.}~\bibnamefont {Stelzer}},\ }\href {\doibase
  10.1007/JHEP06(2011)128} {\bibfield  {journal} {\bibinfo  {journal} {JHEP}\
  }\textbf {\bibinfo {volume} {1106}},\ \bibinfo {pages} {128} (\bibinfo {year}
  {2011})},\ \href {http://arxiv.org/abs/1106.0522}{arXiv:1106.0522
  [hep-ph]}\BibitemShut {NoStop}%
%%CITATION = ARXIV:1106.0522;%%
\bibitem [{\citenamefont {Nikolidakis}\ and\ \citenamefont
  {Smith}(2008)}]{Nikolidakis:2007fc}%
  \BibitemOpen
  \bibfield  {author} {\bibinfo {author} {\bibfnamefont {E.}~\bibnamefont
  {Nikolidakis}}\ and\ \bibinfo {author} {\bibfnamefont {C.}~\bibnamefont
  {Smith}},\ }\href {\doibase 10.1103/PhysRevD.77.015021} {\bibfield  {journal}
  {\bibinfo  {journal} {Phys.Rev.}\ }\textbf {\bibinfo {volume} {D77}},\
  \bibinfo {pages} {015021} (\bibinfo {year} {2008})},\ \href
  {http://arxiv.org/abs/0710.3129}{arXiv:0710.3129 [hep-ph]}\BibitemShut
  {NoStop}%
%%CITATION = ARXIV:0710.3129;%%
\bibitem [{\citenamefont {Chatrchyan}\ \emph
  {et~al.}(2012{\natexlab{b}})\citenamefont {Chatrchyan} \emph
  {et~al.}}]{:2012vh}%
  \BibitemOpen
  \bibfield  {author} {\bibinfo {author} {\bibfnamefont {S.}~\bibnamefont
  {Chatrchyan}} \emph {et~al.} (\bibinfo {collaboration} {CMS Collaboration}),\
  }\href@noop {} {\  (\bibinfo {year} {2012}{\natexlab{b}})},\ \href
  {http://arxiv.org/abs/1212.6194}{arXiv:1212.6194 [hep-ex]}\BibitemShut
  {NoStop}%
%%CITATION = ARXIV:1212.6194;%%
\bibitem [{ATL(2012{\natexlab{b}})}]{ATLAS:2012sna}%
  \BibitemOpen
  \href {http://cds.cern.ch/record/1472674} {\bibfield  {journal} {\bibinfo
  {journal} {ATLAS-CONF-2012-105}\ } (\bibinfo {year} {Aug
  2012}{\natexlab{b}})}\BibitemShut {NoStop}%
\end{thebibliography}%

\end{document}